\documentclass{appolb}

\usepackage{graphicx}
\usepackage{amsmath}
\usepackage[labelsep=period]{caption}
\renewcommand{\thetable}{\Roman{table}}

\providecommand{\keywords}[1]
{
  \small	
  \textbf{\textit{Keywords---}} #1
}
\begin{document}
\title{The weak decay constant of positronium %
\thanks{Based on the presentation of M.P. at the $2^{\text{nd}}$ Symposium on New Trends in Nuclear and Medical Physics, 24 - 26 September 2025, Cracow, Poland.\\
$^a$e-mail: milena.piotrowska@ujk.edu.pl}%
}
\author{M. Piotrowska$^{1,a}$, F. Giacosa$^{1,2}$
\address{$^1$ Institute of Physics, Jan Kochanowski University,\\ \textit{ul. Uniwersytecka 7, 25-406 Kielce, Poland.}\\
$^{2}$ Institute for Theoretical Physics, J. W. Goethe University, \\ \textit{Max-von-Laue-Str. 1, 60438 Frankfurt, Germany.}\\}
}
\maketitle
\begin{abstract}
The positronium, as the lightest purely leptonic bound state, provides an ideal testing ground for quantum field–theoretical (QFT) description of composite systems. While its electromagnetic annihilation is well understood as the dominant decay channel, the weak interaction sector of positronium is strongly suppressed. In this work, we extend the composite QFT framework previously developed for the two–photon decay and introduce the concept of a weak decay constant for para–positronium. This constant, defined in analogy with those of pseudoscalar mesons, quantifies the coupling of the positronium field to the weak axial current and serves as a measure of its internal structure in the electroweak domain. Its numerical value $f_{P}\simeq\frac{m_{e}}{2\sqrt{\pi}}\alpha^{3/2}=89.8593$ \text{ eV} is, as expected, small. Nevertheless, the resulting expression and a related comparison to quarkonium systems allows us to determine the vertex function linking the positronium to its own constituents. 

\end{abstract}
 \textbf{Introductory remarks. }The positronium, composed of an electron and a
positron, represents the simplest neutral bound state described by Quantum
Electrodynamics (QED) \cite{Adkins:2022omi, wheeler, pirene, harris, adkins, czarnecki}. Its theoretical and experimental understanding is
remarkable, reaching precision levels far beyond those attainable for hadronic
systems \cite{Adkins:2022omi, adkins, czarnecki, Kniehl:2000dh, Abreu:2022vei}. In recent years, the positronium has also attracted renewed attention in
applied research, particularly in medical physics. The ability to form
positronium atoms in biological and porous materials enables novel imaging
techniques, and studies performed by the J-PET collaboration have demonstrated
its potential as a diagnostic probe in positron emission tomography and
related fields \cite{Bass:2019ibo, Bass:2023dmv, moskal2019nrp, Harpen:2003zz, moskalexvivo, moskalinvivo}. Nevertheless, positronium remains of interest not only as a
test of QED but also as a conceptual link to composite particles in Quantum
Chromodynamics (QCD). In particular, the para--positronium (p--Ps) is a
pseudoscalar system $(J^{PC}=0^{-+})$ and, in this respect, closely resembles
the neutral pion, the lightest pseudoscalar meson in Quantum Chromodynamics,
as well as to the lightest charm-anticharm state, the $\eta_{c}$ meson. However, this analogy concerns only the quantum numbers and the composite–field structure. While para–positronium can be described within nonrelativistic quantum electrodynamics (NRQED), heavy quarkonia such as the $\eta_{c}$ are treated within the corresponding Quantum Chromodynamics framework (NRQCD), whereas the pion requires a fully relativistic chiral description.

This analogy leads to the discussion of the weak decay constant, widely used
in hadron physics, for positronium states. While the weak decay of the positronium is
very small, this exercise is nevertheless useful because it helps to
understand how to model the coupling of the positronium state to its own
constituents, more specifically the vertex function. As expected, this vertex
function is proportional to the wave function, but a comparison to known
results about quarkonium states allows us to determine this connection. 

More specifically, a composite Quantum Field Theoretical (QFT) model \cite{Weinberg, Hayashi} has been
successfully applied to the two--photon decay of para--positronium, using a
triangle--loop diagram with virtual electrons and a nonlocal vertex function
linked to the bound--state wave function \cite{procsont,Piotrowska:2024nfj}. Such an approach reproduces quite well the
well--known QED decay width and provides a natural connection to similar
processes in mesons. Building upon this formalism, one can proceed one step
further and couple the positronium field to the weak current.

\bigskip

\textbf{Weak decay constants in QCD: brief recall. }In QCD, the pseudoscalar
decay constant $f_{P}^Q$ (where $Q=\bar{q}q$ stands for a generic pseudoscalar quarkonium state) plays a central role. It is defined through the matrix
element
\begin{equation}
\left\langle 0\left\vert J_{W}^{\mu}\right\vert P(p)\right\rangle
=if_{P}^Qp^{\mu}%
\end{equation}
which links the pseudoscalar meson state $\left\vert P(p)\right\rangle $ to
the weak axial--vector current, see e.g. Refs \cite{Godfrey:1985xj,Lucha:1991vn, Azhothkaran:2020ipl}:
\begin{equation}
J_{w}^{\mu}=\bar{\psi}_{q}\gamma^{\mu}\left(  1-\gamma_{5}\right)  \psi
_{q}\text{ .}%
\end{equation}
Here $p^{\mu}$ denotes the four--momentum of the meson, $\psi_{q}$ represents
the quark field, and $\gamma^{\mu}$ and $\gamma_{5}$ are the usual Dirac
matrices. The constant $f_{P}^Q$ characterizes the strength of the coupling
between the meson and the weak current; it determines weak decay rates and
encodes the overlap of the bound--state wave function with the vacuum.

In contrast, the para-positronium, being a purely leptonic system, is governed
almost exclusively by electromagnetic interactions, and its decays are
typically described within the nonrelativistic QED framework \cite{Adkins:2022omi, adkins, czarnecki, Kniehl:2000dh}. Yet, from a
theoretical standpoint, the positronium can also be regarded as a composite
pseudoscalar field. We thus introduce the weak decay constant of positronium,
denoted as $f_{P}$ in the following, by replacing the quark current with the corresponding leptonic
one, $J_{w}^{\mu}=\bar{\psi}_{e}\gamma^{\mu}\left(  1-\gamma_{5}\right)  \psi
_{e}$ \cite{Wolkanowski:2015jtc, Soltysiak, Giacosa:2024scx}.

\bigskip

\begin{figure}
\centerline{%
\includegraphics[width=0.85\linewidth]{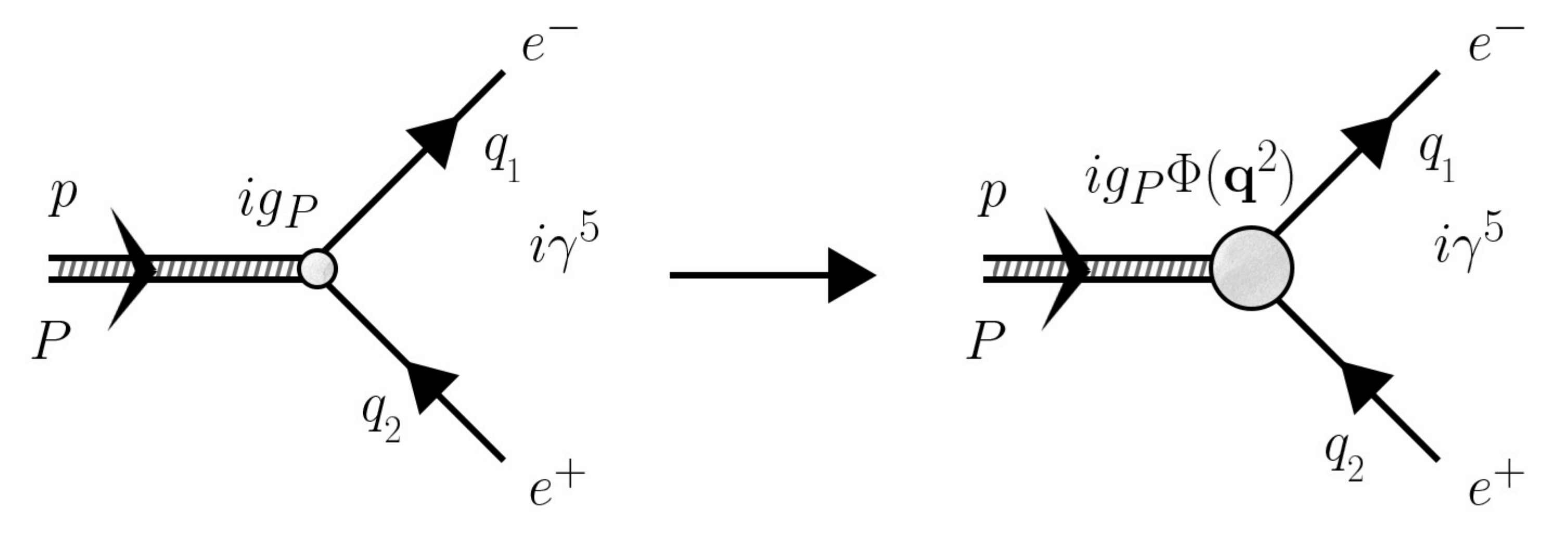}}
\caption{Transition from the local (left) to the nonlocal (right) positronium–electron–positron vertex.}
\label{fig:1}
\end{figure}

\textbf{Composite QFT model for the positronium. }In order to describe the
weak decay constant of positronium, we employ the same composite Quantum Field
Theoretical framework that has been successfully used to study its
electromagnetic annihilation channels \cite{Weinberg, Hayashi, Wolkanowski:2015jtc, Soltysiak, Faessler:2003yf, Giacosa:2004ug, Giacosa:2007bs}. In this approach, the positronium is
treated as an effective pseudoscalar field $P(x)$ representing a bound state
of an electron and a positron. In the local limit, its interaction with the
constituent fermions is described by the effective Lagrangian
\begin{equation}
\mathcal{L}_{P}=g_{P}\,P(x)\,\bar{\psi}_{e}(x)\,i\gamma_{5}\,\psi_{e}(x)
\label{lag}
\end{equation}
where $g_{P}$ is the coupling constant and $\psi_{e}(x)$ denotes the electron
field. The pseudoscalar nature of the positronium state is reflected in the
factor $i\gamma_{5}$, which ensures the correct parity and charge--conjugation
quantum numbers $(J^{PC}=0^{-+})$. Due to the composite nature of the
positronium, the Lagrangian is promoted to be nonlocal  \cite{procsont}. This is
done in analogy to quark-antiquark systems  \cite{Faessler:2003yf, Giacosa:2004ug, Giacosa:2007bs}. In general, the vertex
is modulated by a spacial vertex function $\tilde{\Phi}(x_{1},x_{2},x_{3})$ describing the
smeared nonlocal interaction of the constituents, schematically $\mathcal{L}%
_{P}\rightarrow\mathcal{L}_{P}^{\text{nonlocal}}$. In the rest frame of the
positronium $p=\left(  \sqrt{s},0\right)  $,
\begin{equation}
ig_{P}\rightarrow ig_{P}\Phi(\mathbf{q}^{2})
\end{equation}
with $q=\left(  q_{1}-q_{2}\right)  /2=(0,\mathbf{q})$ and where
$\Phi(\mathbf{q}^{2})$ emerges as the Fourier transform of the vertex
function. See Fig. \ref{fig:1} for a pictorial representation of the passage from the local to the nonlocal case. 
This result can be achieved also in a covariant manner, see details
in \cite{Soltysiak}. The vertex function $\Phi(\mathbf{q}^{2})$ needs to be proportional to
the positronium momentum wave function $A(\mathbf{q}^{2})$. but the question that we aim to discuss here is the precise relation among them. We address this question by using the weak decay constant.

\begin{figure}
\centerline{%
\includegraphics[width=0.58\linewidth]{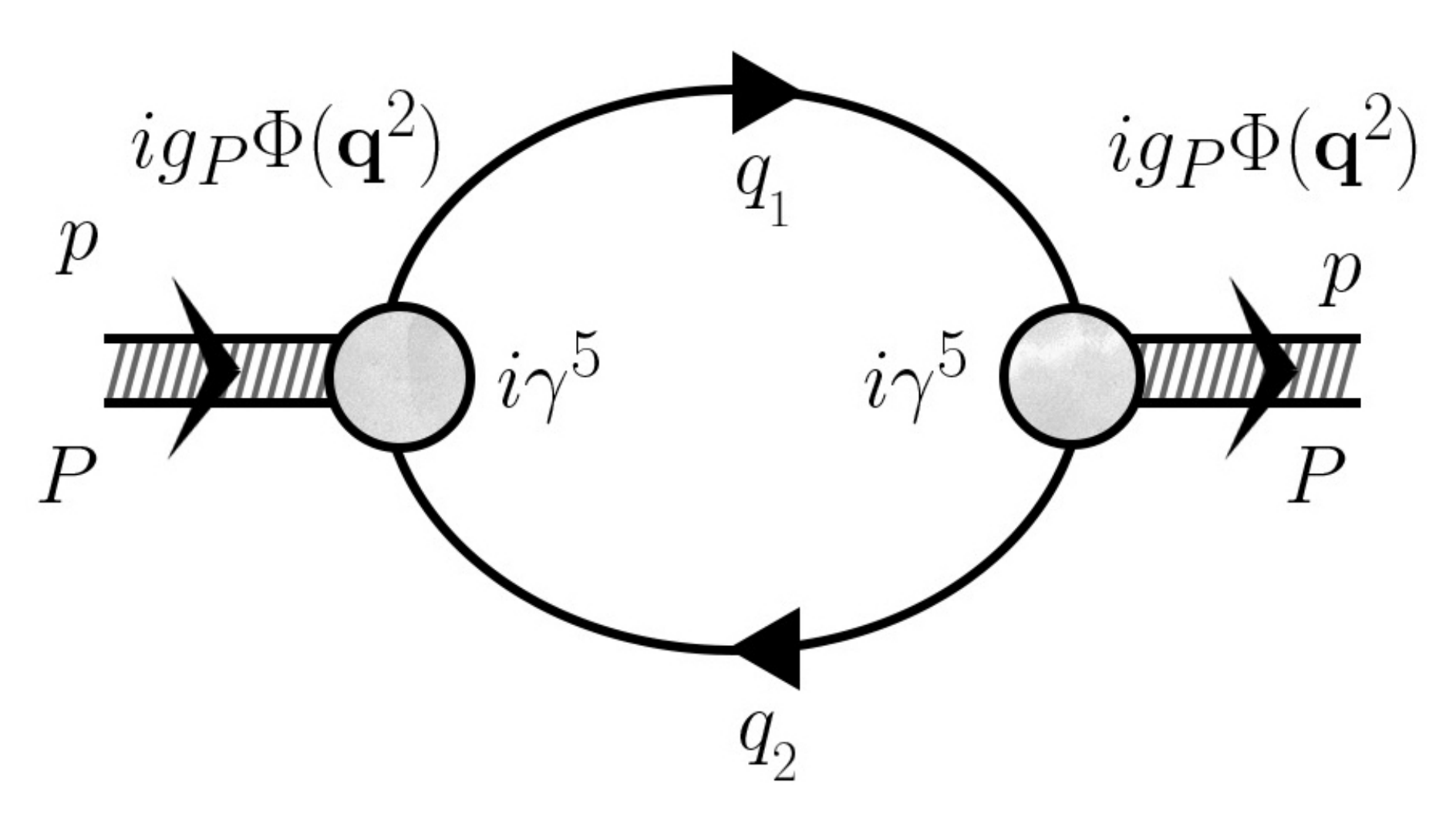}}
\caption{Positronium self–energy diagram with the vertex function $\Phi\left(  \mathbf{q}^{2}\right)$.}
\label{fig:2}
\end{figure}

The coupling constant $g_{P}$ of Eq. (\ref{lag}) is not a free parameter, but is calculated using
the compositeness condition \cite{Weinberg}. For that, the electron-positron loop is
calculated using the Feynman diagrams (see Fig. \ref{fig:2}), leading to (in the rest frame of $P$):
\begin{equation}
\Sigma(s)=i\int\frac{d^{4}q}{(2\pi)^{4}}\frac{Tr\left[  i\gamma^{5}\left(
\not q_{1}+m_{e}\right)  i\gamma^{5}\left(  \not q_{2}+m_{e}\right)  \right]
}{\left(  q_{1}^{2}-m_{e}^{2}+i\varepsilon\right)  \left(  q_{2}^{2}-m_{e}%
^{2}+i\varepsilon\right)  }\Phi^{2}\left(  \mathbf{q}^{2}\right) 
\text{ .}
\end{equation}
Taking into account that
\begin{equation}
Tr\left[  i\gamma^{5}\left(  \not q_{1}+m_{e}\right)  i\gamma^{5}\left(
\not q_{2}+m_{e}\right)  \right]  =4\left(  (q_{1}\cdot q_{2})-m_{e}%
^{2}\right)  \text{ ,}%
\end{equation}
and performing the integral over $q^{0},$ the function $\Sigma(s)$ takes the
compact form%
\begin{equation}
\Sigma(s)=\int\frac{d^{3}q}{(2\pi)^{3}}\frac{8\sqrt{\mathbf{q}^{2}+m_{e}^{2}}%
}{4\left(  \mathbf{q}^{2}+m_{e}^{2}\right)  -s}\Phi^{2}\left(  \mathbf{q}%
^{2}\right)  \text{ .}%
\end{equation}
Hence, its first derivative reads
\begin{equation}
\Sigma^{\prime}(s)=\int\frac{d^{3}q}{(2\pi)^{3}}\frac{8\left(  \sqrt
{\mathbf{q}^{2}+m_{e}^{2}}\right)  }{\left(  4\left(  \mathbf{q}^{2}+m_{e}%
^{2}\right)  -s\right)  ^{2}}\Phi^{2}\left(  \mathbf{q}^{2}\right)  \text{ ,}%
\end{equation}
which allows to determine $g_{P}$ as
\begin{equation}
g_{P}=\frac{1}{\sqrt{\Sigma^{\prime}(s=M_{P}^{2})}}%
\label{gp}
\end{equation}
where $M_{P}=2m_{e}-\frac{m_{e}\alpha^{2}}{4}$ is the parapositronium mass (at
order $\alpha^{2}$).

\bigskip

\begin{figure}
\centerline{%
\includegraphics[width=0.53\linewidth]{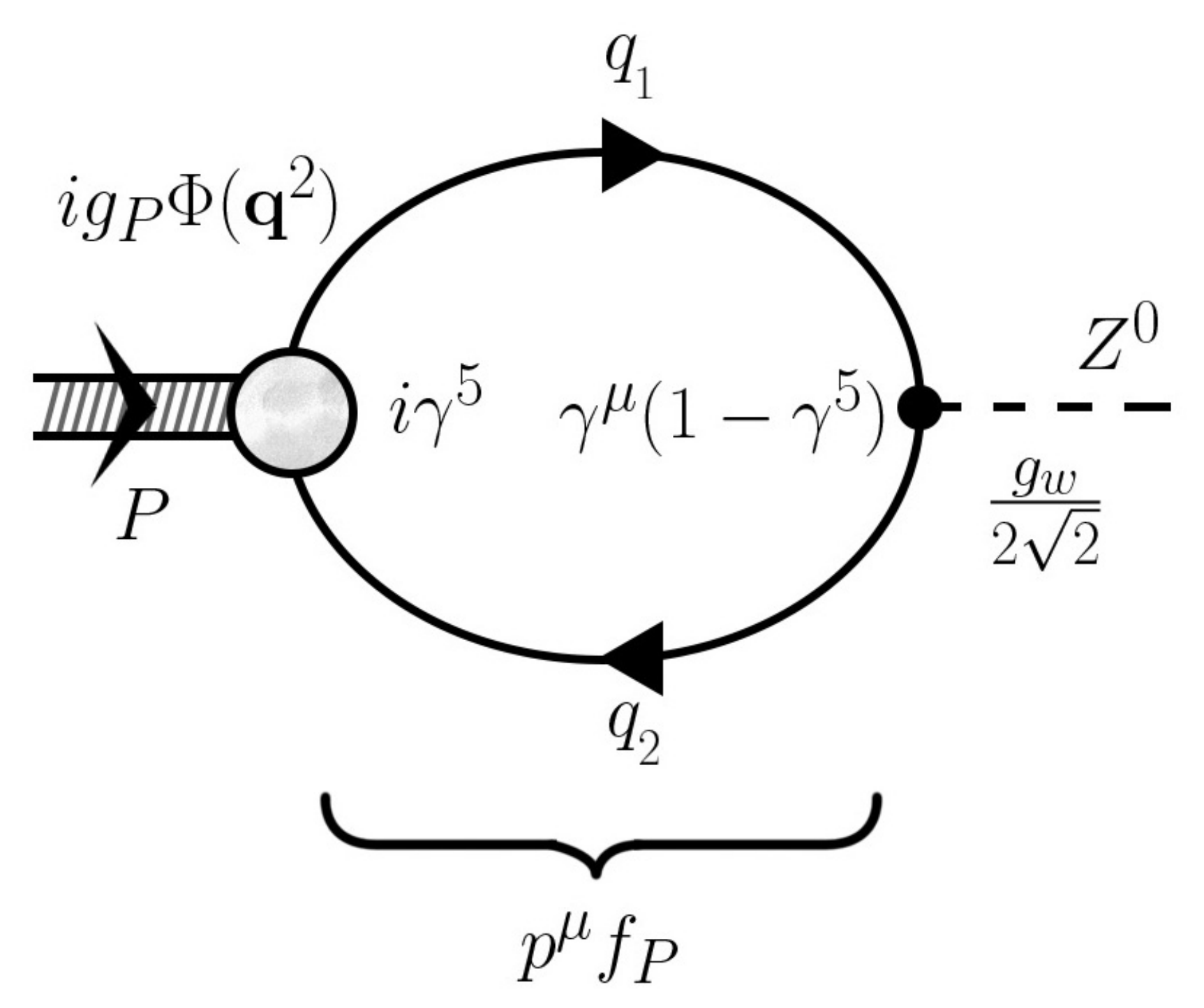}}
\caption{Weak transition $P \rightarrow Z^0$ defining the constant $f_P$.}
\label{fig:3}
\end{figure}

\textbf{Weak decay constant of positronium. }The weak decay constant of the
positronium describes the transition of $P$ into the weak boson $Z^{0}.$ To
this end, we introduce the weak interaction
\begin{equation}
\mathcal{L}_{w}=\frac{g_{w}}{2\sqrt{2}}Z_{\mu}^{0}\left(  \bar{\psi}_{e}%
\gamma^{\mu}(1-\gamma^{5})\psi_{e}+\bar{\psi}_{\nu_{e}}\gamma^{\mu}%
(1-\gamma^{5})\psi_{\nu_{e}}+...\right)
\end{equation}
where $g_{w}$ is the weak coupling constant. Upon considering the Lagrangian $\mathcal{L}%
_{P}^{\text{nonlocal}}+\mathcal{L}_{w},$ the weak decay constant of the field $P$ emerges as (see Fig. \ref{fig:3})
\begin{equation}
p^{\mu}f_{P}=g_{P}\left[  \int\frac{d^{4}q}{(2\pi)^{4}}\frac{Tr\left[
i\gamma^{5}(\not q_{1}+m_{e})\gamma^{\mu}(1-\gamma^{5})(\not q_{2}%
+m_{e})\right]  }{\left(  q_{1}^{2}-m_{e}^{2}+i\varepsilon\right)  \left(
q_{2}^{2}-m_{e}^{2}+i\varepsilon\right)  }\right]  \Phi\left(  \mathbf{q}%
^{2}\right)  \text{ .}%
\end{equation}
It is interesting to note that the same formalism, when the weak vertex
$\gamma^{\mu}(1-\gamma_{5})$ is replaced by the electromagnetic one
$e\gamma^{\mu}( \not  q_{3}-m)^{-1}e\gamma^{\nu}$, reproduces the well--known amplitude for the two--photon
decay of para--positronium. 
The weak and electromagnetic channels thus arise
from a common field--theoretical structure, differing only by the nature of
the external current. This correspondence reinforces the interpretation of
positronium as a convenient theoretical laboratory for exploring the interplay
between composite dynamics and gauge interactions.

Upon calculating the traces and the integral over $q^{0},$ the weak decay
constant of the positronium reads
\begin{equation}
f_{P}=4g_{P}m_{e}\int\frac{d^{3}q}{(2\pi)^{3}}\frac{\Phi\left(  \mathbf{q}%
^{2}\right)  }{\sqrt{\mathbf{q}^{2}+m_{e}^{2}}\left(  4\left(  \mathbf{q}%
^{2}+m_{e}^{2}\right)  -M_{P}^{2}\right)  }.
\end{equation}
This is the main analytical result of this work. In turn, the effective coupling of the positronium to the weak neutral gauge boson reads $\mathcal{L}_{eff}=\frac{g_{w}}{2\sqrt{2}}\ f_{P} \partial _{\mu
}PZ^{0}$, where $g_w$ is the weak coupling.

Next, let us consider a single quark-antiquark state, such as the pseudoscalar
meson $\eta_{c}$. The weak decay constant for this state is expressed as \cite{Ebert:2002qa,Lucha:1991vn,Azhothkaran:2020ipl,Godfrey:1985xj} :
\begin{equation}
f_{P}^{(Q=\eta_{c})}=f_{\eta_{c}}\sim\int\frac{d^{3}q}{(2\pi)^{3}}%
\frac{A_{\eta_{c}}\left(  \mathbf{q}^{2}\right)  }{\sqrt{\mathbf{q}^{2}%
+m_{c}^{2}}}%
\end{equation}
where $A_{\eta_{c}}\left(  \mathbf{q}^{2}\right)  $ is the wave function of
$\eta_{c}$ and $m_{c}$ the (constituent) charm mass. 

A comparison implies the following choice for the positronium vertex function
$\Phi\left(  \mathbf{q}^{2}\right)  $%
\begin{equation}
\Phi\left(  \mathbf{q}^{2}\right)  =\frac{A(\mathbf{q}^{2})}{A(0)}%
\frac{\left(  4\left(  \mathbf{q}^{2}+m_{e}^{2}\right)  -M_{P}^{2}\right)
}{\left(  4m_{e}^{2}-M_{P}^{2}\right)  }%
\end{equation}
where the normalization $\Phi\left(  0\right)  =1$ has been chosen. In turn:
\begin{align}
\Phi\left(  \mathbf{q}^{2}\right)    & =\frac{1}{\left(  1+\frac
{\mathbf{q}^{2}}{\gamma^{2}}\right)  ^{2}}\left(  \frac{\gamma^{2}%
+\mathbf{q}^{2}}{\gamma^{2}}\right)  =\frac{\gamma^{2}}{\gamma^{2}%
+\mathbf{q}^{2}}\text{ ;}\\
\gamma^{2}  & =m_{e}^{2}-\frac{m_{P}^{2}}{4}\text{ .}%
\end{align}
Note, this result agrees with the Ansatz of Ref. \cite{Pestieau:2001ki}. Numerically, the
value of the positronium weak decay constant reads
\begin{equation}
f_{P}=89.4432\text{ eV}%
\end{equation}
to be compared with $f_{\pi}\simeq130$ MeV
for the pion \cite{ParticleDataGroup:2024cfk} or even the 300 MeV value for $\eta_{c}$ \cite{Bedolla:2015mpa}. 

It is also instructive to study the non-relativistic limit.
\begin{equation}
f_{P}\simeq\frac{g_{P}}{\gamma^{2}A(0)}\int\frac{d^{3}q}{(2\pi)^{3}%
}A(\mathbf{q}^{2})=\frac{g_{P}}{\gamma^{2}A(0)}\psi(0).
\end{equation}
On the other hand, taking into account that the derivative of the loop reads%
\begin{align}
\Sigma^{\prime}(m_{P}^{2})  & =\int\frac{d^{3}q}{(2\pi)^{3}}\frac{8\left(
\sqrt{\mathbf{q}^{2}+m_{e}^{2}}\right)  }{\left(  4\left(  \mathbf{q}%
^{2}+m_{e}^{2}\right)  -m_{P}^{2}\right)  ^{2}}\frac{A^{2}(\mathbf{q}^{2}%
)}{A^{2}(0)}\frac{\left(  4\left(  \mathbf{q}^{2}+m_{e}^{2}\right)  -m_{P}%
^{2}\right)  ^{2}}{\left(  4m_{e}^{2}-M_{P}^{2}\right)  ^{2}}\nonumber\\
& =\frac{1}{\left(  4m_{e}^{2}-M_{P}^{2}\right)  ^{2}A^{2}(0)}\int\frac
{d^{3}q}{(2\pi)^{3}}8\left(  \sqrt{\mathbf{q}^{2}+m_{e}^{2}}\right)
A^{2}(\mathbf{q}^{2})\nonumber\\
& \simeq\frac{8m_{e}}{\left(  4m_{e}^{2}-M_{P}^{2}\right)  ^{2}A^{2}(0)}%
\int\frac{d^{3}q}{(2\pi)^{3}}A^{2}(\mathbf{q}^{2})=\frac{8m_{e}}{16\gamma
^{4}A^{2}(0)}%
\end{align}
where the usual normalization $\int\frac{d^{3}q}{(2\pi)^{3}}A^{2}%
(\mathbf{q}^{2})=1$ has been used. Using Eq. (\ref{gp}):
\begin{equation}
g_{P}=\frac{\sqrt{2}\gamma^{2}A(0)}{\sqrt{m_{e}}}=\frac{2\gamma^{2}A(0)}%
{\sqrt{m_{P}}}%
\end{equation}
that leads to the non-relativistic result
\begin{equation}
f_{P}\simeq\frac{2}{\sqrt{m_{P}}}\psi(0)\text{ .}%
\end{equation}
This expression is the Van Royen-Weisskopf formula \cite{VanRoyen:1967nq} for the positronium.
Namely, for a quark-antiquark state, this formula reads
\begin{equation}
f_{P}^{Q}\simeq\frac{\sqrt{N_{c}}2}{\sqrt{m_{Q}}}\Psi_{Q}(0)\text{ }%
\end{equation}
where $N_{c}=3$ is the number of color, $m_{Q}$ the quarkonium mass, and
$\Psi_{Q}(0)$ the quarkonium wave-function evaluated at the origin (for the
large-$N_{c}$ scaling, see the lectures \cite{Giacosa:2024scx}).

For the positronium case, the wave function reads (see e.g. Ref. \cite{Adkins:2022omi}):
\begin{equation}
\psi(r)=\frac{e^{-r/_{a_{P}}}}{\sqrt{\pi}a_{P}^{3/2}}\text{ with }a_{P}%
=\frac{2}{\alpha m_{e}}\text{ ,}%
\end{equation}
leading to %

\begin{equation}
f_{P}\simeq\frac{m_{e}}{2\sqrt{\pi}}\alpha^{3/2}=89.8593\text{ eV,}%
\end{equation}
thus only slightly larger than the previously quoted value. This is expected, since the non-relativistic limit is very accurate for the positronium. 

\bigskip

\textbf{Conclusions. }In this work, the weak decay constant of positronium has
been defined and evaluated within the framework of a composite Quantum Field
Theoretical model. The positronium field, treated as a pseudoscalar bound
state of an electron and a positron, was coupled to the weak axial current in
analogy with the treatment of pseudoscalar mesons in QCD. The resulting
expression for $f_{P}$ is strongly suppressed, scaling as $f_{P}\simeq
\alpha^{3/2}\,m_{e}/(2\sqrt{\pi})$.

Although such a quantity is strongly suppressed in positronium phenomenology,
it provides a consistent formal bridge between leptonic and hadronic composite
systems. In turn, the momentum vertex function is connected to the wave function as $\phi(\mathbf{q}^2) \sim A(\mathbf{q})(\mathbf{q}^2+m_e^2-m_P^2)$. This result may be used in future studies linking positronia and quarkonia. 

The approach developed here may also serve also as a starting point for
further studies of parity--violating effects and excited positronium states
within the same QFT framework. Moreover, even if extremely small, the process
may offer a transition to particles beyond the SM, such as the putative
$X(17)$ state \cite{Krasznahorkay:2015iga}. The effective interaction is of the type $c_{X}f_{P}\partial_{\mu} P X^{\mu}$, where the dimensionless constant $c_{X}$ describes the coupling of $X(17)$ to electrons. Notably, this interaction implies that $X(17)$ is an axial-vector state and is reminiscent of the $a_1(1260)$--$\pi$ mixing in effective hadronic models, e.g. \cite{Giacosa:2024epf, Ko:1994en}.
Interestingly, the composite model may also be used to test the case in which the $X(17)$ is a pseudoscalar object. In this case, the effective interaction is $c_X \lambda_{PX} PX$, where $\lambda_{PX}$ is the $PX$ transition loop. Both models are worth studying in the future.

\bigskip

\textbf{Acknowledgements:} This work was supported by the Minister of Science (Poland) under the `Regional Excellence Initiative' program (project no.: RID/SP/0015/2024/01, with sub-projects RID/2024/LIDER/08 and RID/2025/LIDER/02). 


\end{document}